\documentclass{preprint}
\usepackage{graphics}
\begin{document}

{\ }
\vspace{7cm}
\begin{center}
This is an unedited preprint. The original publication is available at \\ {\ } \\
http://www.springerlink.com \\ {\ } \\
http://www.doi.org/10.1023/A:1017056831002
\end{center}
\newpage

\begin{frontmatter}  
%
\title{High-spin low-spin transition}       
\author[A]{H. Gr\"{u}nsteudel}     
\author[B]{H. Paulsen}
\author[B]{H. Winkler}
\author[B]{A. X. Trautwein\thanks{Corresponding author}}
\author[A]{H. Toftlund}
\address[A]{Department of Chemistry, University of Odense, DK-5230 Odense,
Denmark}   
\address[B]{Institut f\"ur Physik, Medizinische Universit\"at L\"ubeck,
D-23538 L\"ubeck, Germany 
                  \email{trautwein@physik.mu-luebeck.de} }
\runningauthor{H. Gr\"{u}nsteudel et al.}
\runningtitle{High-spin low-spin transition}
%
\begin{abstract}  
Temperature dependent nuclear inelastic-scattering (NIS) of synchrotron radiation was applied to investigate both spin states of the spin-crossover complex [Fe(tpa)(NCS)$_2$] (tpa=tris(2-pyridylmethyl)amine).
A remarkable increase of the iron-ligand bond stretching upon spin crossover has unambiguously been identified by comparing the measured NIS spectra of with theoretical simulations based on density-functional calculations.
\end{abstract}
%
\end{frontmatter}

\begin{figure}
\hspace{\fill}
\scalebox{0.6}{\includegraphics{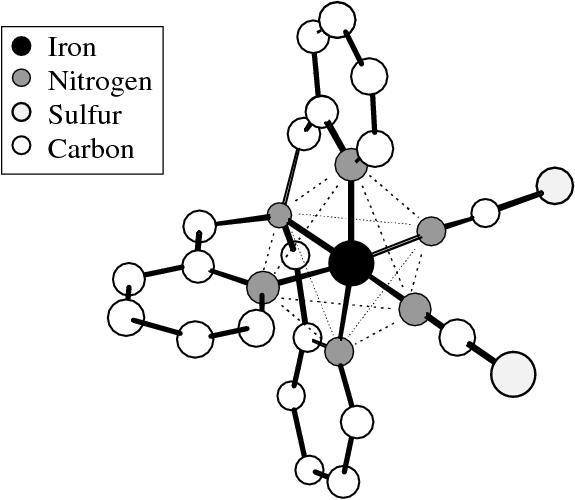}}
\hspace{\fill}
\caption{\label{fig1} Geometry of the HS isomer of [Fe(tpa)(NCS)$_2$]
optimized with the density functional method B3LYP\cite{Becke93} together with the split valence basis set 3-21G*\cite{Binkley80}
(Hydrogens omitted).} 
\end{figure}

The iron(II) complex [Fe(tpa)(NCS)$_2$] (Fig. \ref{fig1}, tpa =
tris(2-pyridylmethyl)\linebreak[4] amine) belongs to the family of thermally
driven spin-crossover complexes, which exhibit a transition from a low-spin
(LS) to a high-spin (HS) state by increasing the temperature. These complexes
are very promising materials for optical information storage
\cite{Guetlich94}. 

IR measurements on several spin-crossover complexes with a central [FeN$_6$]
octahedron indicate a remarkable increase of the Fe-N bond stretching
frequencies from about 25 to 30\,meV in the HS state to about 50 to 60\,meV in
the LS state\cite{Koenig91}. In this case nuclear inelastic scattering (NIS)
turns out a very valuable alternative to IR and Raman spectroscopy because the
Fe-N stretching modes can be definitely identified in the NIS spectra, whereas
the IR and Raman spectra are rather complex in this frequency region
\cite{Takemoto73} making an unambiguous assignment of these modes very
difficult. 

NIS spectra were recorded at the Nuclear Resonance Beamline ID 18 of the
European Synchrotron Radiation Facility (ESRF) in Grenoble, France
\cite{Rueffer96}. The 6\,GeV electron storage ring was operated in 16 bunch mode
and the purity of the filling (population of parasitic bunches compared to the
single bunch) was better than $10^{-9}$. The incident beam was monochromatized
by a double-crystal Si(111) premonochromator to the bandwidth of 2.5\,eV at the
energy of 14.413\,keV. A further decrease of bandwidth down to 6 meV was
obtained with a "nested" high-resolution monochromator \cite{Ishikawa92}. The
sample was mounted in a closed-cycle cryostat to allow measurements at
different temperatures. An avalanche photo-diode with $10\times 10$ mm$^2$
area, 100 $\mu$m active thickness, and a time resolution of less than 1 ns has
been used as a detector \cite{Baron95} to count the 14.413\,keV $\gamma$-quanta
and, mainly, the K-fluorescence photons ($\approx 6.4$ keV). The data were
collected during several energy scans with 140 steps on average, each with 2
meV stepsize and 10\,s measuring time. All individual scans were corrected for
the approximately 9\%\ decrease of beam intensity of the storage ring during
the 1500\,s required for each scan and added up afterwards. 

\begin{figure}
\hspace{\fill} \scalebox{0.6}{\includegraphics{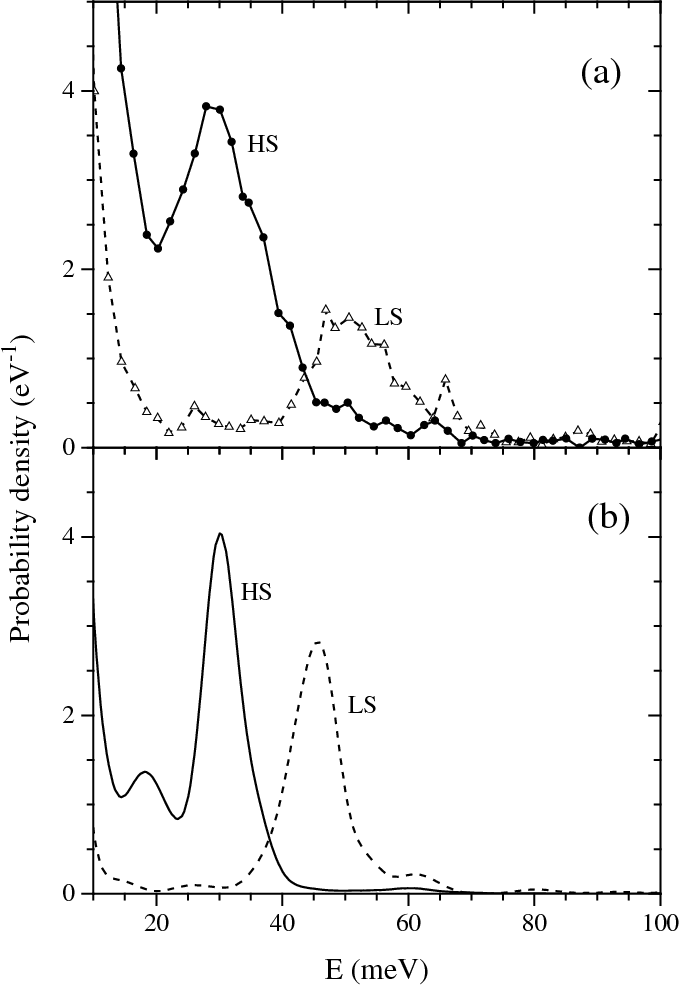}} \hspace{\fill}
\caption{\label{fig2} Measured (a) and simulated (b) NIS spectra of the HS (solid lines, $\bullet$) and LS (dashed lines, $\triangle$)
isomers of [Fe(tpa)(NCS)$_2$].}
\end{figure}

The NIS spectra of the HS and LS isomers of [Fe(tpa)(NCS)$_2$] exhibit central
peaks of 12 and 7\,meV linewidth, respectively, and a pronounced inelastic peak
at 30 meV in the HS state and at 50\,meV in the LS state
(Fig. \ref{fig2})\cite{Gruensteudel98a,Gruensteudel98,Paulsen98}. Comparing the
intensity of the pronounced inelastic peaks in the HS and the LS spectrum it
should be kept in mind, that the HS peak at 30 meV is located on the shoulder
of the corresponding central peak. The linewidth of the LS peak observed at 50
meV is significantly larger than the linewidth of the corresponding central
peak. The LS peak should, therefore, be regarded as a superposition of two or
more individual peaks. The LS spectrum exhibits another, rather small peak at
66 meV, which is invisible in the HS spectrum. 

The first momentum $\overline{E}$ of the measured absorption probability
density $S_{\rm meas}(E)$ of the HS isomer (1.8\,meV) is in reasonable
agreement with the recoil energy $E_R$\,=\,1.96\,meV of the free $^{57}$Fe
nucleus, 
as expected according to Lipkin's rule, \cite{Lipkin60} whereas the first
momentum of the LS spectrum amounts to 4\,meV, which is about twice as large as
$E_R$. This large measured $\overline{E}$ is due to the increased attenuation
of the incident radiation at nuclear resonance \cite{Sturhahn95}. This
phenomenon can be neglected if the elastic peak has a small weight $[f_{\rm
LM}^{\rm HS}=0.21 (1)]$. If however a large fraction of the spectrum belongs to
the elastic peak $[f_{\rm LM}^{\rm LS}=0.68 (1)]$ the measured absorption
probability density has to be corrected. For this reason the NIS spectrum of
the LS isomer was normalized according to 
\begin{equation}
  S(E) = S_{\rm meas}(E) \frac{E_R}{\overline{E}} + \left( 1 -
\frac{E_R}{\overline{E}} \right) R(E) 
\end{equation}
The resolution function $R(E)$ describes the energy distribution of the
incident radiation and is assumed to be an even function. 

For the HS and the LS isomers of [Fe(tpa)(NCS)$_2$] electronic structure
calculations were performed using the density functional theory (DFT) method
B3LYP, \cite{Becke93} implemented in the {\sc gaussian94} program system
\cite{Frisch95} together with the split valence 3-21G* basis set
\cite{Binkley80}. The geometries were fully optimized applying the Berny
algorithm to redundant internal coordinates \cite{Peng96}. Force constants were
calculated analytically at the same level of theory using the optimized
geometries, and the resulting vibrational frequencies were corrected by the
scaling factor 0.9613 as has been proposed by Wong \cite{Wong96} for the
6-31G* basis set. The calculated normal modes for both isomers have been used
to simulate the absorption probability density $S(E)$ according to the
procedure described in Ref\cite{Paulsen98}. 

No X-ray structures are available that can be compared with the calculated geometries, but the calculated bond lengths of the HS and LS isomers qualitatively resemble the increase of the Fe-N bond distances upon spin crossover of about 10 to 20\,pm observed in various spin-crossover complexes with a central [FeN$_6$] octahedron \cite{Koenig91}.

The vibrational spectra of the HS and the LS isomers of [Fe(tpa)(NCS)$_2$] consist of 135 normal modes and are, in the following discussion, subdivided into a high-frequency region above 75\,meV and a low-frequency region below 75\,meV.
The high-frequency region is of minor interest for the purposes of this study, since the vibrational modes in this frequency region do not contribute to the mean-square displacement (msd) of the iron nucleus and, therefore, can not be observed by NIS. 

Among the 41 normal modes of the low-frequency region the iron-ligand bond stretching vibrations are of special interest here.
Due to the almost octahedral environment of the iron center three out of six Fe-N stretching modes are invisible in NIS and IR spectra.
Those modes that transform according to the A$_1$ and E$_{\rm g}$ representations of the ideal octahedron do not contribute to the msd of the iron nucleus or to the variation of the electric dipole moment.
Only the remaining three modes, that transform according to the T$_{\rm 1u}$ representations can be observed in NIS and IR spectra.
These three modes, with calculated frequencies of 29.9, 30.1, and 35.3\,meV for the HS state and 42.8, 46.6, and 52.6\,meV for the LS state, give rise to prominent peaks in the simulated NIS spectra of both isomers of [Fe(tpa)(NCS)$_2$].
Considerable contributions to the calculated absoption probability also arise from N-Fe-N bending modes in the range from 3 to 20\,meV.
These modes can not be identified in the experimental spectra because they are superimposed by the much larger contributions to the NIS spectra originating from the acoustical phonons.

By IR spectroscopy \cite{Hojland83} Fe-N bond stretching frequencies of 59.5
and 66.0\,meV have been found for the LS isomer, while for the region below
35\,meV, that is difficult to reach, no frequencies are reported. The Fe-N bond
stretching frequencies calculated for the LS isomer are about 12.4\,meV
smaller than the IR values given above; however, they are in good agreement
with the frequencies obtained from NIS. The broad peak at 50\,meV observed in
the measured NIS spectrum of the LS isomer (Fig.\ref{fig2}) represents the
envelope of the three Fe-N stretching modes in the range of 45 to 55\,meV.
The pronounced peak at 30\,meV in the NIS spectrum of the HS isomer is
assigned to 
the same modes (Fig.\ref{fig2}). These modes reflect, according to the
intensity of the peaks, the substantial contributions to the msd of the iron
nucleus that is associated with the three T$_{\rm 1u}$ Fe-N stretching modes. 

According to the normal mode analysis the low-intensity peak at 66\,meV in the
measured NIS spectrum as well as the line at 65.7\,meV in the IR spectrum must
be assigned to a mode which has predominantly N-C-S bending character and to
some extent Fe-N stretching character. The mixed character of this mode is due
to interactions between Fe-N stretching and N-C-S bending modes, which are
close in energy in the LS isomer. 

The calculated N-C-S bending modes of the HS isomer do not show any admixture
of Fe-N stretching modes because of the relatively large energy gap of about
30\,meV between these modes. Correspondingly the NIS spectrum of the HS isomer
does not exhibit a peak at the respective energy. In summary, the NIS spectra
of the LS isomer as well as the DFT calculations suggest, that the IR line
attributed previously to an Fe-N bond stretching mode of the LS isomer should
be assigned to a bending mode of the NCS group instead. As a result the
frequency shift of the Fe-N stretching mode upon spin crossover is about 40\%\
smaller than assumed earlier. DFT calculations for another spin-crossover
complex with NCS groups, i.e., [Fe(phen)$_2$(NCS)$_2$] (phen =
1,10-phenanthroline), lead to a similar conclusion. 

The measured Lamb-M\"{o}ssbauer factor of [Fe(tpa)(NCS)$_2$] is decreasing
from $f_{\rm LM}^{\rm LS}=0.68 (1)$ for the LS state at 34\,K to $f_{\rm
LM}^{\rm HS}=0.21 (1)$ for the HS state at 200\,K
\cite{Gruensteudel98a}. Comparison of these values with the calculated
molecular Lamb-M\"{o}ssbauer factors ($f_{\rm mol}^{\rm LS}=92$ and $f_{\rm
mol}^{\rm HS}=0.52$) indicates, that for both spin states the major part of
the iron msd is due to {\it inter}-molecular vibrations. However, the msd of
the HS state contains, according to the calculations, also significant
contributions from {\it intra}-molecular vibrations. 

Due to its ability to focus on few modes out of a rather complex vibrational
spectrum NIS can be a complementary or, for certain problems, even a superior
alternative to conventional methods like IR and Raman spectroscopy. A good
example is the investigation of iron(II) spin-crossover complexes as presented
here. IR and Raman spectra are rather complex in the frequency range of the
Fe-N bond stretching modes (20 - 60\,meV). Even if the isotope technique is
used, the assignment of these modes to the observed bands often remains
doubtful as has been demonstrated for [Fe(tpa)(NCS)$_2$]. In the NIS spectra,
however, the Fe-N stretching modes could be unambiguously identified.

\acknowledgements

The authors acknowledge the support by A.I. Chumakov, R. R\"{u}ffer, and
H. F. Gr\"{u}nsteudel and by the relevant ESRF services during these
measurements, and the financial support by the European Union
(ERB-FMRX-CT-0199) via the TMR-TOSS-network, by the German Research Foundation
(DFG) and by the German Federal Ministery for Education, Science, Research and
Technology (BMBF).


%
\end{document}